\documentclass[proof]{pasj00}

\begin{document}
\SetRunningHead{TAMAGAWA ET AL.}{Suzaku Observations of Tycho's SNR}
\Received{2008/05/02}
\Accepted{2008/05/19}

\title{Suzaku Observations of Tycho's Supernova Remnant}

\author{Toru~\textsc{Tamagawa},\altaffilmark{1,2}
        Asami~\textsc{Hayato},\altaffilmark{1,2}
        Satoshi~\textsc{Nakamura},\altaffilmark{1,2}
        Yukikatsu~\textsc{Terada},\altaffilmark{3}
	Aya~\textsc{Bamba},\altaffilmark{4}
        Junko~S.~\textsc{Hiraga},\altaffilmark{1}
	John~P.~\textsc{Hughes},\altaffilmark{5}
	Una~\textsc{Hwang},\altaffilmark{6}
	Jun~\textsc{Kataoka},\altaffilmark{7}
	Kenzo~\textsc{Kinugasa},\altaffilmark{8}
	Hideyo~\textsc{Kunieda},\altaffilmark{9}
	Takaaki~\textsc{Tanaka},\altaffilmark{4}
	Hiroshi~\textsc{Tsunemi},\altaffilmark{10}
	Masaru~\textsc{Ueno},\altaffilmark{7}
	Stephen~S.~\textsc{Holt},\altaffilmark{11}
	Motohide~\textsc{Kokubun},\altaffilmark{4}
	Emi~\textsc{Miyata},\altaffilmark{10}
	Andrew~\textsc{Szymkowiak},\altaffilmark{12}
	Tadayuki~\textsc{Takahashi},\altaffilmark{4}
	Keisuke~\textsc{Tamura},\altaffilmark{4}
	Daisuke~\textsc{Ueno},\altaffilmark{9}
	\\and
	Kazuo~\textsc{Makishima}\altaffilmark{13,1}
}
\altaffiltext{1}{RIKEN, 2-1 Hirosawa, Wako, Saitama 351-0198}
\email{tamagawa@riken.jp}
\altaffiltext{2}{Department of Physics, Tokyo University of Science, 1-3
Kagurazaka, Shinjyuku-ku, Tokyo 162-8601}
\altaffiltext{3}{Department of Physics, Saitama University, Sakura,
Saitama 338-8570}
\altaffiltext{4}{Institute of Space and Astronautical Science, Japan
Aerospace Exploration Agency, 3-1-1 Yoshinodai, Sagamihara, Kanagawa
229-8510}
\altaffiltext{5}{Department of Physics and Astronomy, Rutgers
University, 136 Frelinghuysen Road, Piscataway, NJ 08854-8019, USA}
\altaffiltext{6}{NASA Goddard Space Flight Center, Greenbelt, MD 20771,
USA}
\altaffiltext{7}{Department of Physics, Faculty of Science, Tokyo
Institute of Technology, 2-12-1, Meguro-ku, Ohokayama, Tokyo 152-8551}
\altaffiltext{8}{Gunma Astronomical Observatory, 6860-86 Nakayama,
Takayama-mura, Agatsuma-gun, Gunma 377-0702}
\altaffiltext{9}{Department of Astrophysics, Nagoya University,
Furo-cho, Chikusa-ku, Nagoya 464-8063}
\altaffiltext{10}{Department of Earth and Space Science, Graduate School
of Science, Osaka University, 1-1 Machikaneyama, Toyonaka, Osaka
560-0043}
\altaffiltext{11}{F. W. Olin College of Engineering, Needham, MA 02492,
USA}
\altaffiltext{12}{Department of Physics, Yale University, New Haven, CT
06520, USA}
\altaffiltext{13}{Department of Physics, The University of Tokyo, 7-3-1
Hongo, Bunkyo-ku, Tokyo 113-0033}

%


\maketitle

\begin{abstract}
 Tycho's supernova remnant was observed by the XIS and HXD instruments
 onboard the Suzaku satellite on 2006 June 26--29 for 92~ks. The
 spectrum up to 30~keV was well fitted with a two-component model,
 consisting of a power-law with photon index of 2.7 and a thermal
 bremsstrahlung model with temperature of 4.7~keV. The former component
 can alternatively be modeled as synchrotron emission from a population
 of relativistic electrons with an estimated roll-off energy of around
 1~keV. In the XIS spectra, in addition to the prominent Fe K$_\alpha$
 line (6.445~keV), we observe for the first time significant K$_\alpha$
 line emission from the trace species Cr and Mn at energies of 5.48~keV
 and 5.95~keV, respectively. Faint K$_\beta$ lines from Ca (4.56~keV)
 and Fe (7.11~keV) are also seen. The ionization states of Cr and Mn,
 based on their line centroids, are estimated to be similar to that of
 Fe K$_\alpha$ (Fe XV or XVI).
\end{abstract}

 \section{Introduction}
 
 Extensive studies at X-ray and TeV energies have been carried out in
 recent years to reveal the origin of Galactic cosmic rays
 (CRs). Diffusive shock acceleration at the blast waves of supernova
 remnants (SNRs) is widely believed to be a prime source for producing
 CRs up to energies of the so-called ``knee'' in the CR spectrum around
 1000~TeV. The ASCA observation of SN 1006 \citep{koyama1995} revealed a
 featureless power-law component, interpreted as originating from
 synchrotron emission near the outer shell and appearing well separated
 from a thermal component filling the projected interior. Such a clear
 situation is, however, not usually the case with the other youngest
 SNRs, in which any nonthermal synchrotron X-ray emission from the blast
 wave is difficult to separate from the relatively more dominant thermal
 X-ray components. To more readily achieve this separation, we focus on
 the higher energy X-ray band where the thermal contribution is low.

 One of the brightest remnants among the Galactic historical supernovae
 is Tycho's SNR (observed by Tycho Brahe in AD 1572). Synchrotron
 emission, like that from SN~1006, has been expected from this remnant
 based on shock acceleration models (see, for example,
 \cite{ammosov1994}), at high enough X-ray brightness levels to offer
 the possibility of observationally separating the nonthermal from
 thermal emission. The first evidence for hard X-ray emission in Tycho's
 SNR was provided by the A-2 experiment on HEAO 1 \citep{pravdo1979}
 which measured the integrated spectrum of the entire remnant up to
 25~keV. Using the 4.5--20~keV Ginga spectrum of the SNR,
 \citet{fink1994} detected a power-law component with photon index
 $\alpha_p=2.72$, in combination with a thermal bremsstrahlung continuum
 and an Fe K$_\alpha$ emission line. The RXTE satellite detected the
 hard continuum spectrum up to 20~keV and measured a photon index
 $\alpha_p\sim$3 \citep{petre1999}.

 Recent Chandra and XMM-Newton observations of Tycho's SNR have provided
 spatially resolved spectra up to photon energies around 10~keV. One
 remarkable feature of the high angular resolution data was the
 discovery of spectrally-featureless, geometrically-thin filaments
 encircling nearly the entire outer rim of the SNR
 \citep{hwang2002}. Although the featureless nature of the spectra from
 the outer rim can be accommodated by the either power-law or thermal
 bremsstrahlung models, various lines of evidence favor a nontherml
 synchrotron origin for the emission \citep{warren2005}. Detailed models
 for the X-ray spectra and morphology of these filaments
 \citep{cassam2007} strongly indicate that the outer rim X-ray emission
 of this SNR comes via synchrotron radiation from electrons
 shock-accelerated up to energies of a few 10~TeV. Similar features have
 now been seen in other young remnants of historical supernovae
 \citep{bamba2005}.

 Tycho's SNR also shows intense atomic emission lines which were studied
 in detail with Tenma by \citet{tsunemi1986}, who showed for the first
 time that the K$_\alpha$ line energies of Si, S, Ar, Ca, and Fe were
 significantly lower than those expected under ionization equilibrium
 conditions, requiring that the X-ray emitting plasma be in a strongly
 under-ionized and therefore nonequilibrium condition. Investigations of
 Tycho's thermal emission, especially its spatial variation, were also
 carried out by ASCA \citep{hwang1997}, Chandra \citep{hwang2002}, and
 XMM-Newton \citep{decourchelle2001}. The most recent detailed results
 on the ejecta properties of the Tycho's SNR come from
 \citet{badenes2006}, who compared the integrated X-ray spectrum to
 theoretical models for the ejecta X-ray emission of thermonuclear
 (i.e., Type Ia) supernovae.

 In the present paper, we study the recent broad-band X-ray observation
 of Tycho's SNR by the Suzaku satellite. Utilizing the long exposure
 duration and the excellent CCD spectral resolution we conduct a
 sensitive search for faint emission lines. We also use the broad
 spectral coverage of Suzaku (the remnant is detected from photon
 energies of 0.5~keV up to a few tens of keV) to separate the nonthermal
 component from the dominant thermal X-ray component.

 \section{Observations and Data Reduction}
 
  \subsection{Observations}
  
  Launched on 2005 July 10 as a joint Japan-US mission
  \citep{mitsuda2007}, Suzaku carries four X-ray telescopes (XRT;
  \cite{serlemitsos2007}) each of which illuminates an X-ray imaging
  spectrometer (XIS; \cite{koyama2007}) with
  \timeform{18'}$\times$\timeform{18'} field of view. The payload
  includes a co-aligned non-imaging hard X-ray detector (HXD;
  \cite{takahashi2007}) as well. While the entire HXD and the high
  energy portion of the XIS spectra play a major role in studying the
  nonthermal emission, the lower energy part of the XIS data a also
  useful for accurately constraining the thermal component.

  Tycho's SNR was observed by Suzaku on 2006 June 26 through June 29 as
  part of the mission's Science Working Group program. The observation
  was carried out with the remnant's center placed at the center of the
  XIS CCD chip; the pointing coordinates were ($\alpha$,
  $\delta$)=(\timeform{00h25m20s}, \timeform{64D08m18s}) (equinox
  J2000.0). A nearby sky region, at coordinates ($\alpha$,
  $\delta$)=(\timeform{00h36m54s}, \timeform{64D17m42s}),
  \timeform{1D.27} offset from the SNR along the Galactic plane, was
  observed on 2006 June 29 in order to acquire background data. There
  are no known bright X-ray sources in the HXD field-of-view at the
  offset observation.

  During these observations, 16 out of 64 silicon PIN diodes of the HXD
  were operated at a lower bias voltage (400~V) to avoid unexpected
  breakdown of the sensors. The other PIN diodes were operated at the
  nominal bias voltage of 500~V. The four XIS sensors, three front
  illuminated CCD chips (XIS 0, XIS 2 and XIS 3) and a back illuminated
  one (XIS 1), were operated in the standard mode.

  \subsection{HXD-PIN Data Reduction}

  In the analysis, we excluded the data from the 16 PIN diodes operated
  at the lower bias voltage, because it is difficult at the present time
  to estimate their energy response and background accurately enough. We
  used cleaned events from the revision 1.2 data processing of the HXD
  data reduction software. Data were excluded if acquired during
  potential high background parts of the orbit (where the cutoff
  rigidity less than 8~GeV/c or for 500~s after passage through the
  South Atlantic Anomaly), or when the target elevation was less than
  5$^{\circ}$ above the horizon. The data selection was done using the
  {\tt XSELECT} package of {\tt HEASOFT} version 6.2. The net exposure
  after data selection was 92.0~ks and 45.8~ks for the Tycho and offset
  observations, respectively.

  Although a generic non--X-ray background model is provided by the
  detector team, we used the offset observation data to estimate the
  total instrument and sky background of HXD-PIN. This is because we
  expect to achieve higher accuracy for the HXD-PIN background
  subtraction when using the offset observation, which was carried out
  for a relatively long time just after the on-source observation, under
  nearly the same conditions for the non--X-ray background. We
  superposed orbital phases of the on-source and offset observations,
  and extracted the events from exactly the same orbital phase
  regions. The net on-source and off-source exposures used for making
  spectra are 74.5~ks and 37.3~ks, respectively.

\begin{figure}
   \begin{center}
    \FigureFile(120mm,120mm){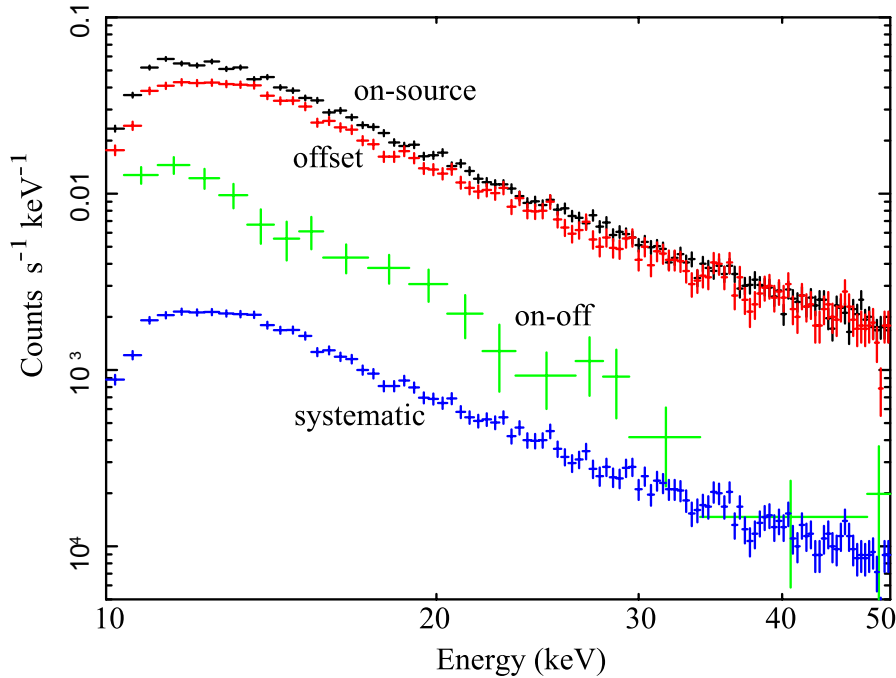}
   \end{center}
   \caption{On-source (black), off-source (red), and
   background subtracted (green; on-off) spectra of Tycho's SNR, taken
   with HXD-PIN. The blue points show 5\% of the background spectrum, as
   a measure of the systematic error associated with the background
   subtraction.}  \label{fig:onset_offset}
  \end{figure}

  Figure \ref{fig:onset_offset} shows the on-source, offset and
  background subtracted HXD-PIN spectra from 10 through 50~keV. We
  estimated the systematic error in the background spectra from Earth
  occultation data, which gives a measure of the non--X-ray
  background. We selected data with elevation from the Earth rim $\le
  -5^{\circ}$ for both the on-source and offset observations, retaining
  identical values for the other selection criteria. We compared the
  occultation events between the two observations, and found that they
  agree with each other to within 5\% accuracy. In the following we
  adopt a value of 5\% for the systematic error in the spectrum. To
  evaluate the background-subtraction accuracy in greater detail, we
  made a background subtracted 12--30~keV light curve and fitted it with
  a constant. The fit yielded a mean count rate and the associated 90\%
  confidence-level error of (5.66$\pm$0.31)$\times$10$^{-2}$ cts
  s$^{-1}$, with $\chi^2$/dof=1691/1547. The light curve is thus
  consistent with a constant source (as expected for the SNR) thereby
  validating the background subtraction technique. In figure
  \ref{fig:onset_offset}, the level of background systematic error
  (calculated as 5\% of the offset spectrum) is shown. We have thus
  securely detected the hard X-ray emission from Tycho's SNR up to
  30~keV. The lowest end of the PIN spectrum extends to $\sim$12~keV, as
  determined by thermal noise in the diodes \citep{kokubun2007}.

  For an extended source like Tycho's SNR, an additional correction
  factor must be applied to the HXD-PIN effective area when trying to
  estimate the source flux. The remnant is about \timeform{8'.2} in
  diameter, and the XIS pointed right at the center of the remnant
  (``XIS nominal position''). Although this correction factor should be
  included in the ancially response files (ARFs) based on the specific
  morphology of the source, the current software tool {\tt hxdarfgen}
  can calculate the response only for point sources. Therefore, to mimic
  a ring-like morphology (i.e., a rim-brightened structure as expected
  for Tycho's SNR), we assumed 8 point sources placed uniformly around a
  circle of \timeform{4'.1} radius centered on the pointing
  direction. Then, we calculated an ARF for each of them, taking into
  account the \timeform{3'.5} offset between the XIS and HXD optical
  axes. The final ARF was obtained by just averaging the eight ARFs. To
  evaluate systematic errors associated with this ARF, we also generated
  a disk-like ARF with the same radius. The difference between the
  annular and disk-like ARFs was confirmed to be less than 1\% in the
  12--30 keV band.

  \subsection{XIS Data Reduction}

\begin{figure}
   \begin{center}
    \FigureFile(160mm,160mm){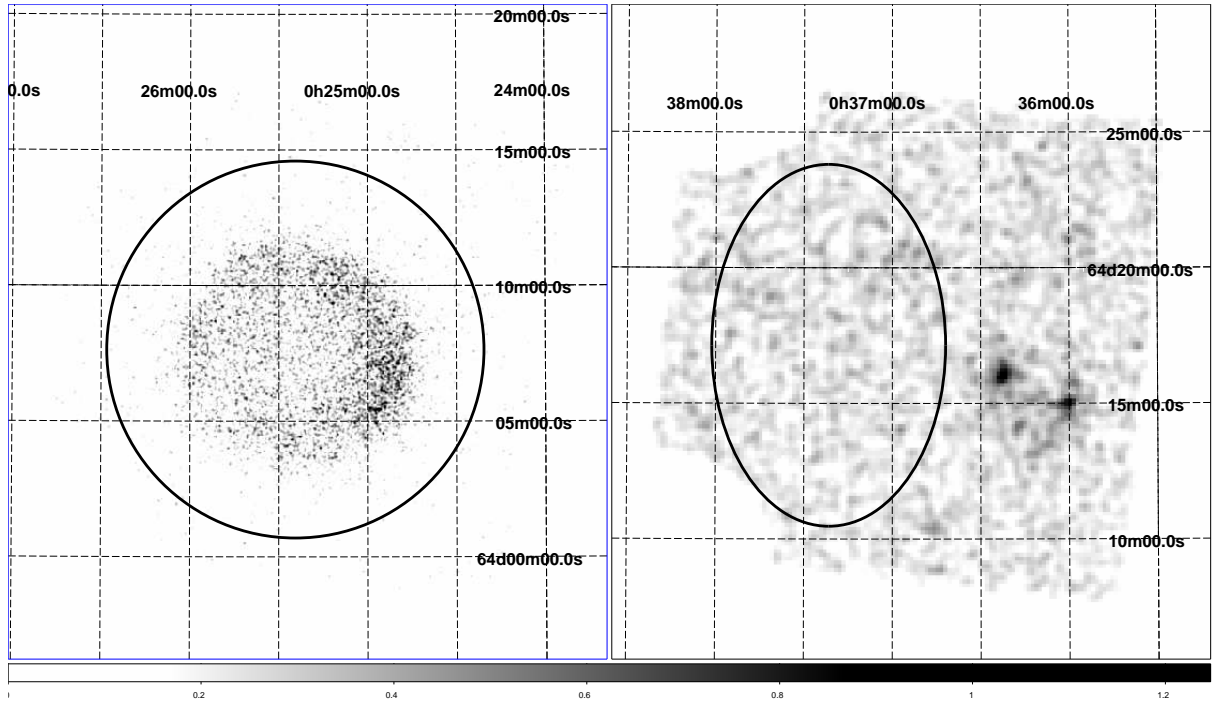}
   \end{center}
   \caption{(Left) An X-ray image of Tycho's SNR taken by XIS0 in
   $>$7~keV. The circle on the image indicates the region from which the
   on-source spectrum was derived. (Right) A 5--12 keV image of the
   offset region on XIS0. The ellipse indicate the region employed to
   produce the background spectrum.}  \label{fig:xis0_image}
  \end{figure}

  For the XIS data reduction, we used cleaned events (revision 1.2) of
  the three front-illuminated CCDs (XIS 0, XIS 2 and XIS 3). The left
  panel of figure \ref{fig:xis0_image} shows the XIS image at energies
  between 7 and 12 keV. The region we employed to extract the on-source
  spectrum is shown in the figure. The right panel of figure
  \ref{fig:xis0_image} shows the image of the offset observation, where
  an ellipse on the image indicates the region from which we extracted
  the background spectrum. Note that this region excluded an unknown
  diffuse, faint object visible in the XIS image of the offset
  observation. To evaluate the XIS background systematic error, we made
  another background file from the Lockman Hole observation (80~ks
  exposure). The difference between the offset and Lockman Hole
  background spectra is 1\% in the integrated 5--12~keV count rate and
  the spectral shape is identical within statistical errors, we have
  decided to use the offset background in the analysis.

  ARF files describing the XIS effective area, combined with the X-ray
  mirror response, were generated by {\tt xissimarfgen}
  \citep{ishisaki2007}. To make the XIS ARF consistent with that of PIN,
  we generated the XIS ARFs assuming an annular source with a radius of
  \timeform{4'.1}. Since we interested mainly in the hard X-ray
  emission, this assumption is justified by the XIS image above 7~keV
  which reveals a clear ring-line source morphology.

 \section{Analysis and Results}
 
  \subsection{Characterization of HXD-PIN spectrum}

  Figure \ref{fig:pin_spec} shows the background subtracted PIN spectrum
  of Tycho's SNR in the 12--30~keV band. We fitted the data successfully
  with a power-law model, and obtained a photon index of
  $\alpha_p=2.83^{+0.56}_{-0.53}$(stat)$^{+0.56}_{-0.03}$(syst) with
  $\chi^2$/dof=25.5/21. The statistical error is shown at a confidence
  level of 90\%, while the systematic error refers to the 5\%
  uncertainty in estimating the background using the offset data. The
  residuals from the fit are also given in figure \ref{fig:pin_spec}. If
  we assume the source of the emission to have an annular shape with a
  radius of \timeform{4'.1}, the 12--30~keV model flux is
  ($1.90^{+0.13}_{-1.28}\;^{+0.48}_{-0.30}$)$\times$10$^{-11}$
  ergs~cm$^{-2}$~s$^{-1}$. Alternatively, we fitted the PIN data with a
  thermal bremsstrahlung spectrum, and obtained a temperature of
  $kT=12.4^{+8.2}_{-3.9}\;^{+0.9}_{-2.1}$ keV with
  $\chi^2$/dof=27.1/21. The 12--30~keV model flux is
  ($1.88^{+0.13}_{-0.89}\;^{+0.52}_{-0.62}$)$\times$10$^{-11}$
  ergs~cm$^{-2}$~s$^{-1}$. Thus, the two fits are both acceptable,
  although the power-law model is statistically somewhat favored.

  \subsection{Emission lines above 4 keV}

\begin{figure}
   \begin{center}
    \FigureFile(130mm,130mm){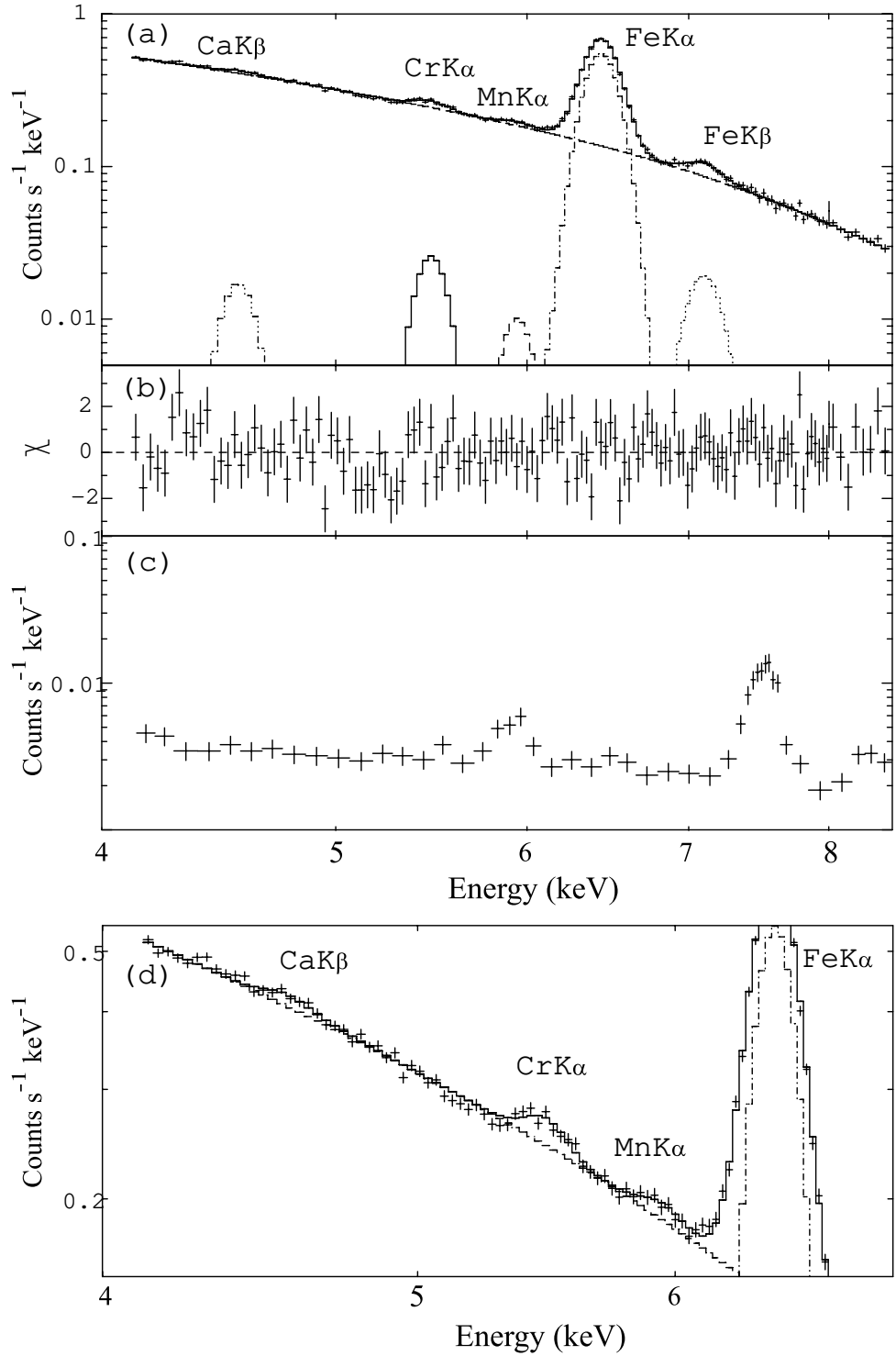}
   \end{center}
   \caption{(a) A summed spectrum of XIS0, 1, and 2 above 4~keV. A
   power-law and five Gaussian models are fitted to the data. (b)
   Residuals of the fitting. (c) A background spectrum. (d) A magnified
   spectrum in energies of 4--6.5~keV. }  \label{fig:lines}
  \end{figure}

  The atomic emission lines below 4~keV have been well studied by
  previous missions, including ASCA \citep{hwang1997}, Chandra, and
  XMM-Newton \citep{decourchelle2001, badenes2006}. Here we focus on the
  spectra above 4~keV, where Suzaku has a larger effective area and a
  lower background level than other missions. Figure \ref{fig:lines}a
  shows the XIS spectrum in the 4.1--8.5~keV energy range, where we
  observe an intense Fe K$_\alpha$ emission line. We fitted the spectrum
  with a composite model consisting of a power-law and several
  Gaussians. The best fit parameters were $\alpha_p=2.7$ and the Fe
  K$_\alpha$ line center energy was 6.445~keV. The measured line center
  energy is in an approximate agreement with that obtained with Tenma,
  $6.40\pm 0.03$~keV \citep{tsunemi1986}. This fit was not statistically
  acceptable ($\chi^2$/dof=447.6/180), and there were clear residuals
  around 4.5, 5.5, 5.9 and 7.1~keV. The fit was significantly improved
  ($\chi^2$/dof=209.1/170) by the addition of four more Gaussian
  lines. The best fit model is also shown in figure \ref{fig:lines}a,
  and the residuals are shown in figure \ref{fig:lines}b. Figure
  \ref{fig:lines}d shows a magnified view of the spectrum around the
  faint lines. The best fit parameters of the four Gaussians, including
  the central energy, flux and equivalent width, are summarized in the
  Table \ref{table:lines}. The background spectrum subtracted from the
  source spectrum is shown in figure \ref{fig:lines}c. The weak line at
  $\sim$5.90~keV is leakage from the on-board $^{55}$Fe calibration
  source and only appears in sensor XIS 0. The line at $\sim$7.47~keV is
  nickel fluorescence from the nearby structure of the XIS camera.

\begin{figure}
   \begin{center}
    \FigureFile(140mm,140mm){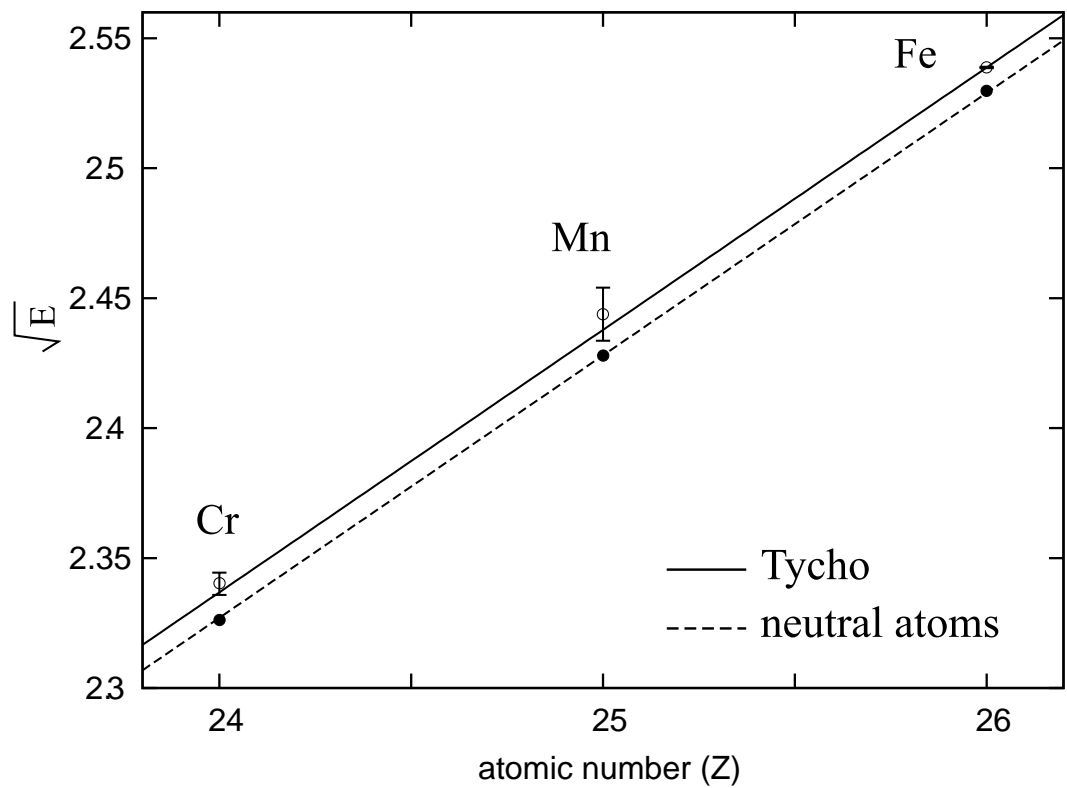}
   \end{center}
   \caption{Measured line energies composed with a prediction of the
   Moseley's law. The energies of K$_{\alpha}$ lines of neutral atoms
   from Cr to Fe are plotted on the same figure.}  \label{fig:moseley2}
  \end{figure}

  The lines at 4.56, 5.48, 5.95, and 7.11~keV have not been reported
  previously. Given the significant flux from He-like Ca K$_\alpha$
  \citep{badenes2006}, the simplest explanation for the 4.56~keV line is
  He-like Ca K$_\beta$. (Furthermore, the expected He-like Ca K$_\beta$
  energy centroid of 4.58~keV is consistent with our line measurements.) 
  The next two lines, 5.48 and 5.95~keV, are probably K$_\alpha$ lines
  from elements near Fe on the periodic table. They exhibit slightly
  higher energies than the K$_\alpha$ lines of neutral Cr and Mn, which
  are centered at 5.41, and 5.90~keV, respectively. To evaluate their
  nature, we plotted their energy $E_i$ as a function of atomic number,
  $Z_i$, and fitted to Moseley's law \citep{moseley1913};
  \begin{eqnarray}
   \label{eq:moseley}
   E_i = a \; (Z_i - b)^2.
  \end{eqnarray}
  It is known that the normalization factor is
  $a=3/4\;Ry=1.02\times10^{-2}$~keV with Rydberg energy ($Ry$) and the
  screening factor is $b\sim1$ for K$_\alpha$ lines of neutral
  atoms. Figure \ref{fig:moseley2} shows the line energies determined in
  this analysis and K$_\alpha$ lines from neutral atoms of {\it
  Z}=24--26, together with results obtained by fitting equation
  (\ref{eq:moseley}). In the fits, we fixed $a$ to that of neutral
  species, and allowed $b$ to vary. The best fit value obtained was
  $b=0.862\pm0.002$, and the new K$_\alpha$ lines fell on Moseley's law
  within uncertainties. We thus conclude that the lines are from
  similarly ionized states of Cr, Mn, and Fe. We note that there is a
  weak contamination line at 5.90~keV (figure \ref{fig:lines}c) but this
  appears only in XIS 0, while the 5.95~keV line was detected in all XIS
  instruments.

  If we extrapolate our best fit to Moseley's law, we estimate the
  centroid energy of the Co ($Z=27$) and Ni ($Z=28$) K$_\alpha$ lines to
  be 6.97 and 7.51~keV. Neither value matches the line centroid of the
  7.11~keV feature, from which we conclude that the line at 7.11~keV is
  probably the Fe K$_\beta$ blend. The low central energies of both the
  Fe K$_\alpha$ and Fe K$_\beta$ line blends indicate that the emitting
  iron is in an extremely low ionization condition. The central energies
  of K$_\alpha$ and K$_\beta$ are consistent with those of Fe XV or XVI
  as calculated by \citet{mendoza2004}. Assuming these ionization
  states, the Fe K$_\beta$ to K$_\alpha$ flux ratio,
  $f$(K$_\beta$)/$f$(K$_\alpha$)=0.04$\pm$0.01, is consistent with
  \citet{mendoza2004} who calculated this flux ratio for K-vacancy
  configurations of iron atoms.

  Although there was no significant excess emission at energies
  corresponding to the Ni K$_\alpha$ in the XIS spectrum (figure
  \ref{fig:lines}a), we estimated the upper limit of the Ni K$_\alpha$
  flux by adding a Gaussian with its centroid energy fixed at
  7.51~keV. The fits yielded a 90\%-confidence upper limit flux and
  equivalent width of $8.15\times10^{-6}$~ph~cm$^{-2}$~s$^{-1}$ and
  25.1~eV, respectively.

  \subsection{PIN and XIS joint fitting with conventional models}
  \label{subsect:broadband_spectra}

  To characterize the continuum spectrum of Tycho's SNR, we fitted the
  XIS spectra above 5~keV simultaneously with the 12--30~keV PIN
  spectrum. Since the lines from Cr and Mn introduced above are not
  taken into account in current nonequilibulium ionization models, we
  fitted the data with conventional (power-law or thermal
  bremsstrahlung) models plus four Gaussians. In the fitting, the low
  energy absorption was fixed to $N_{\rm H}=0.7\times10^{22}$~cm$^{-2}$
  as obtained by \citet{cassam2007}, while the other parameters were
  left free. As calibrated by \citet{kokubun2007} using the Crab Nebula,
  the model normalization to fit the PIN data was set 13\% higher than
  that of the XIS data.

\begin{figure}
   \begin{center}
    \FigureFile(100mm,100mm){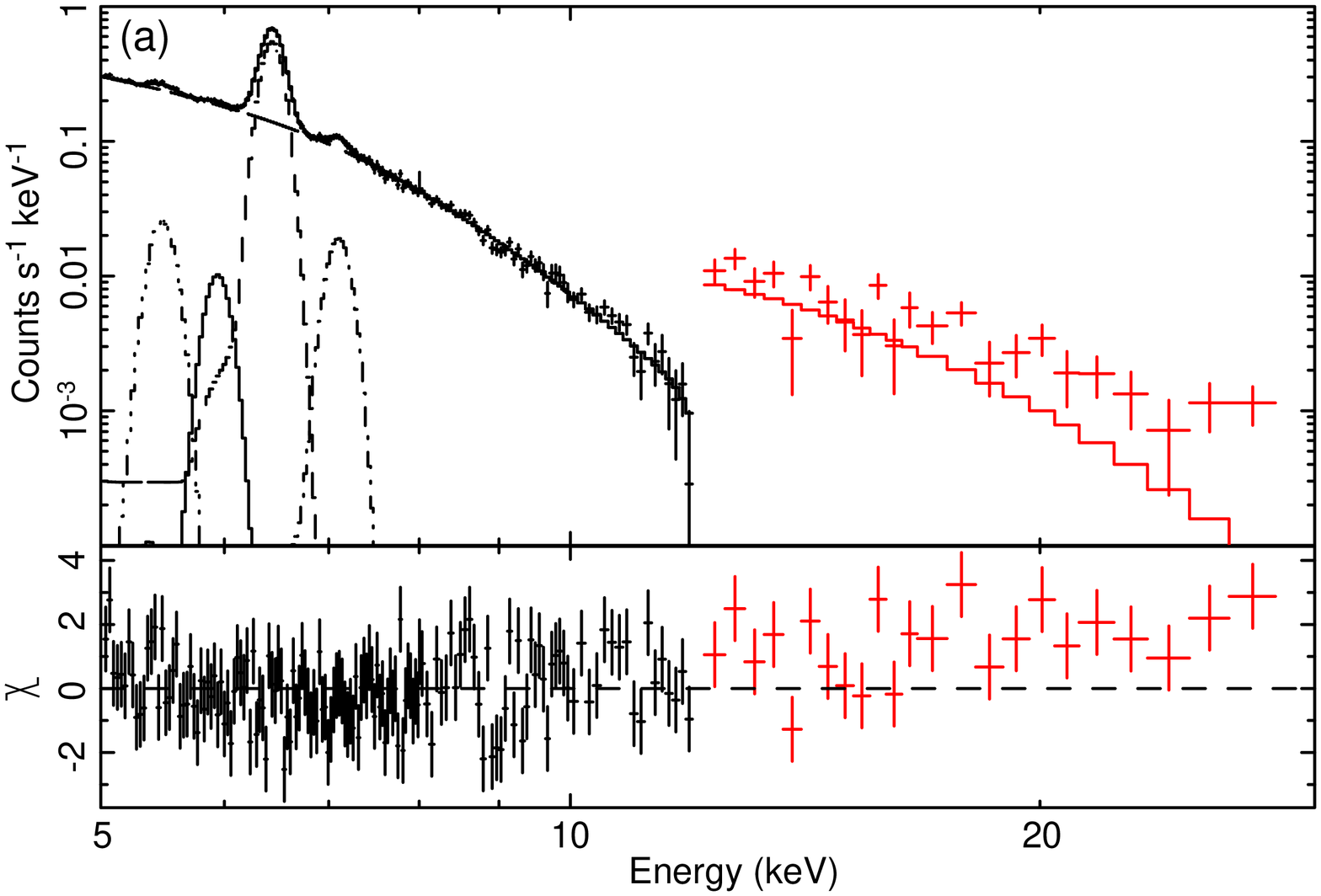}
    \FigureFile(100mm,100mm){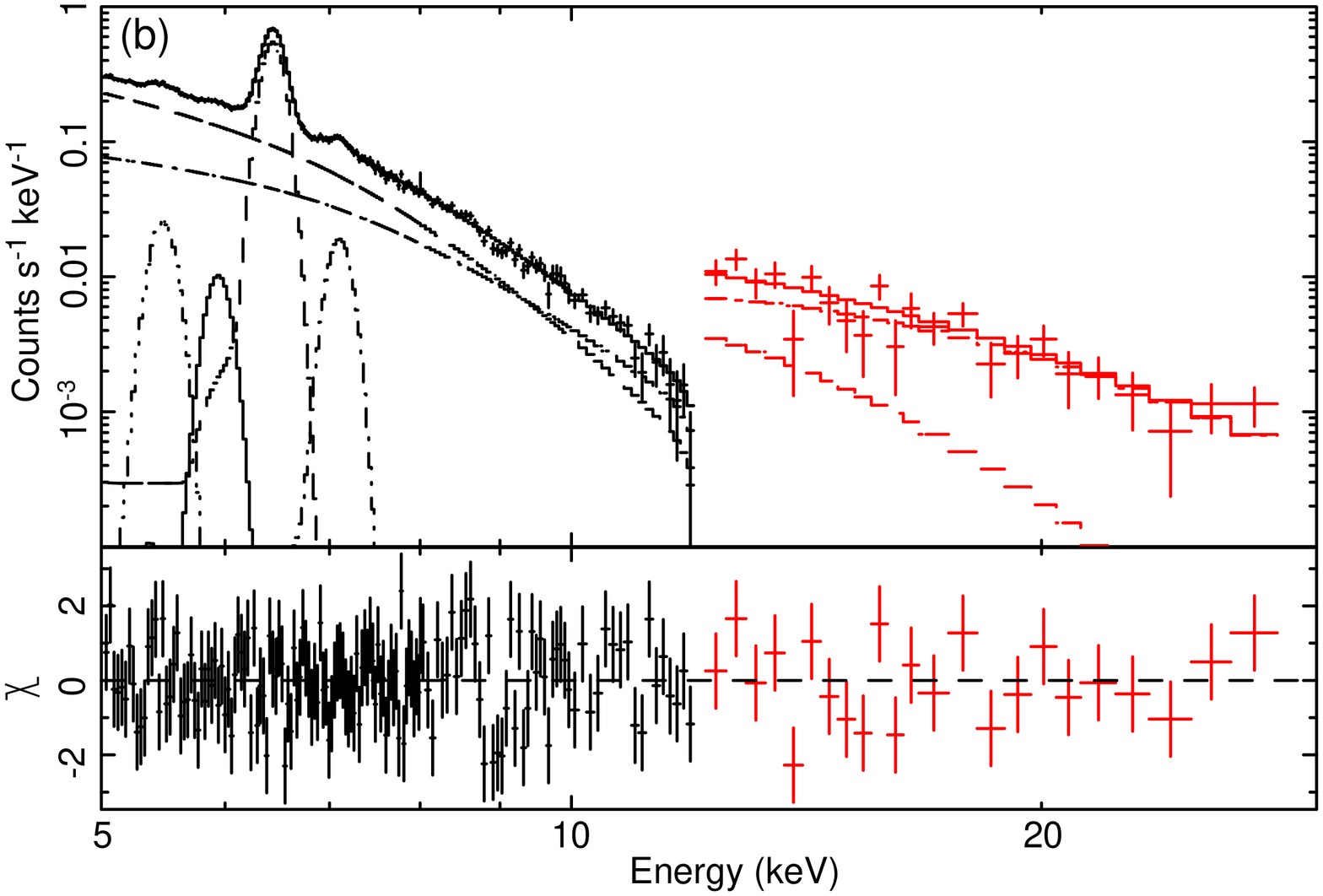}
    \FigureFile(100mm,100mm){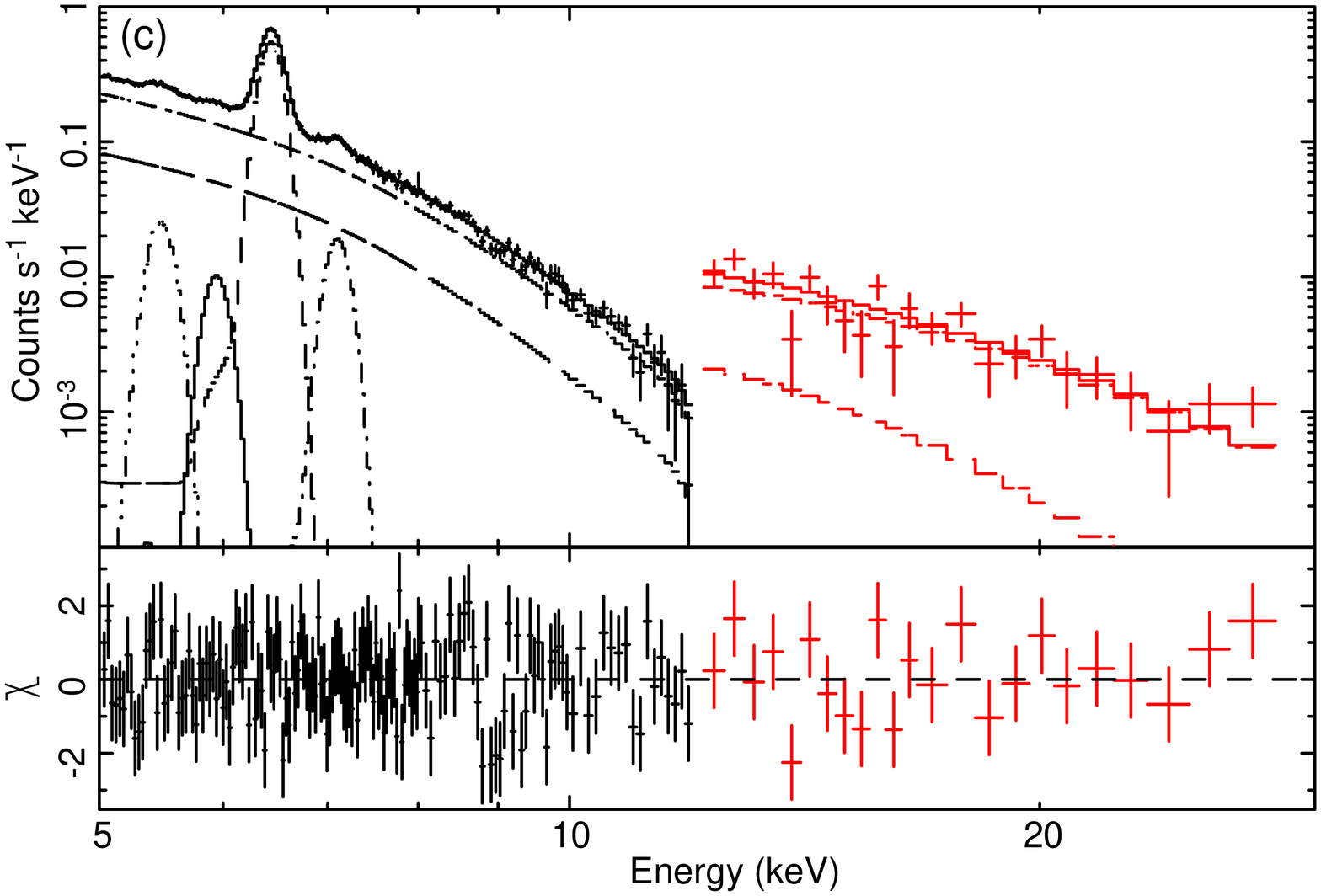}
   \end{center}
   \caption{XIS and PIN spectra fitted with (a) a thermal bremsstrahlung
   model, (b) two thermal bremsstrahlung models, and (c) a thermal
   bremsstrahlung and a power-law models. Fit residuals are also plotted
   in the figures. The emission lines of Cr, Mn, and Fe are represented
   in the fit by Gaussian models.}  \label{fig:xis_pin_fit}
  \end{figure}

  First, we fitted the spectra with a thermal bremsstrahlung model plus
  Gaussians. Figure \ref{fig:xis_pin_fit}a shows the fit result and
  residuals. The best-fit parameter value was
  $kT=5.14^{+0.14}_{-0.12}$~keV. The fit yielded a relatively large
  $\chi^2$/dof=253.8/175, and there was clear excess above 10~keV. In an
  attempt to reduce the $\chi^2$ value, we next added another
  bremsstrahlung component to the model. Figure \ref{fig:xis_pin_fit}b
  shows the best fit spectrum and residuals. The best fit value was
  $kT_{\rm low}=3.48^{+0.59}_{-1.19}$~keV and $kT_{\rm
  high}=22.9^{+\infty}_{-12.9}$~keV with $\chi^2$/dof=185.9/173. These
  values are similar to those from the Ginga spectral analysis, in which
  the temperatures were $kT_{\rm low}=2.71$~keV and $kT_{\rm
  high}=11.6$~keV \citep{fink1994}. Although the two thermal
  bremsstrahlung model is statistically acceptable, the temperature of
  the hotter component, $kT_{\rm high}$, is probably too high to reflect
  thermal emission from the remnant. At face value, it suggests a shock
  speed of a few thousand km~s$^{-1}$, but recent work indicates that
  very little electron heating occurs in such high speed shocks, since
  the level of electron heating is observed to decrease strongly with
  increasing shock speed \citep{rakowski2003}. The actual shock velocity
  corresponding to a 22.9~keV electron temperature in Tycho's SNR would
  have to be so high as to be unreasonable. Moreover, the upper limit on
  the higher temperature component is not constrained as shown in
  figure~\ref{fig:xispin_brbr_cont}. This strongly suggests that the
  harder spectral component is of nonthermal origin, rather than
  thermal.
  
  In our ultimate model, we fitted the spectrum with a thermal
  bremsstrahlung component and a
  power-law. Figure~\ref{fig:xis_pin_fit}c shows the model spectra and
  residuals. The best fit parameters were $kT=4.71^{+0.66}_{-1.02}$~keV
  and $\alpha_p=2.69^{+0.23}_{-1.27}$ with $\chi^2$/dof=185.9/173. The
  power-law model flux at 1~keV is
  7.3$\times$10$^{-2}$~ph~cm$^{-2}$~s$^{-1}$~keV$^{-1}$, which
  corresponds to the 10--20~keV band model flux of 1.3$\times$10$^{-11}$
  ergs~cm$^{-2}$~s$^{-1}$. From these results, we conclude that the
  power-law component with $\alpha_p=2.69$ extends beyond 20~keV, which
  is the uppermost energy of the Ginga observations \citep{fink1994}, up
  to at least 30~keV.

 \section{Discussion}
 \label{sect:discuss}

 \subsection{Origin of the hard X-ray emission}

 We have found that the hard power-law tail of Tycho's SNR's spectrum
 extends up to energies of at least 30~keV with a photon index of
 $\sim$2.69. The photon index is consistent with those observed by
 Chandra from the rim regions of the SNR in 0.5--10~keV low energy band
 \citep{hwang2002,warren2005,cassam2007}. We further verified that the
 overall normalization of the power-law component in the integrated
 Chandra spectrum is consistent with the power-law normalization we
 found here. Thus, we conclude that the featureless power-law spectra
 observed in the Chandra energy region naturally extends to the Suzaku
 HXD-PIN energy region up to at least 30~keV.

 \subsection{Roll-off frequency of the synchrotron spectrum}

\begin{figure}
   \begin{center}
    \FigureFile(100mm,100mm){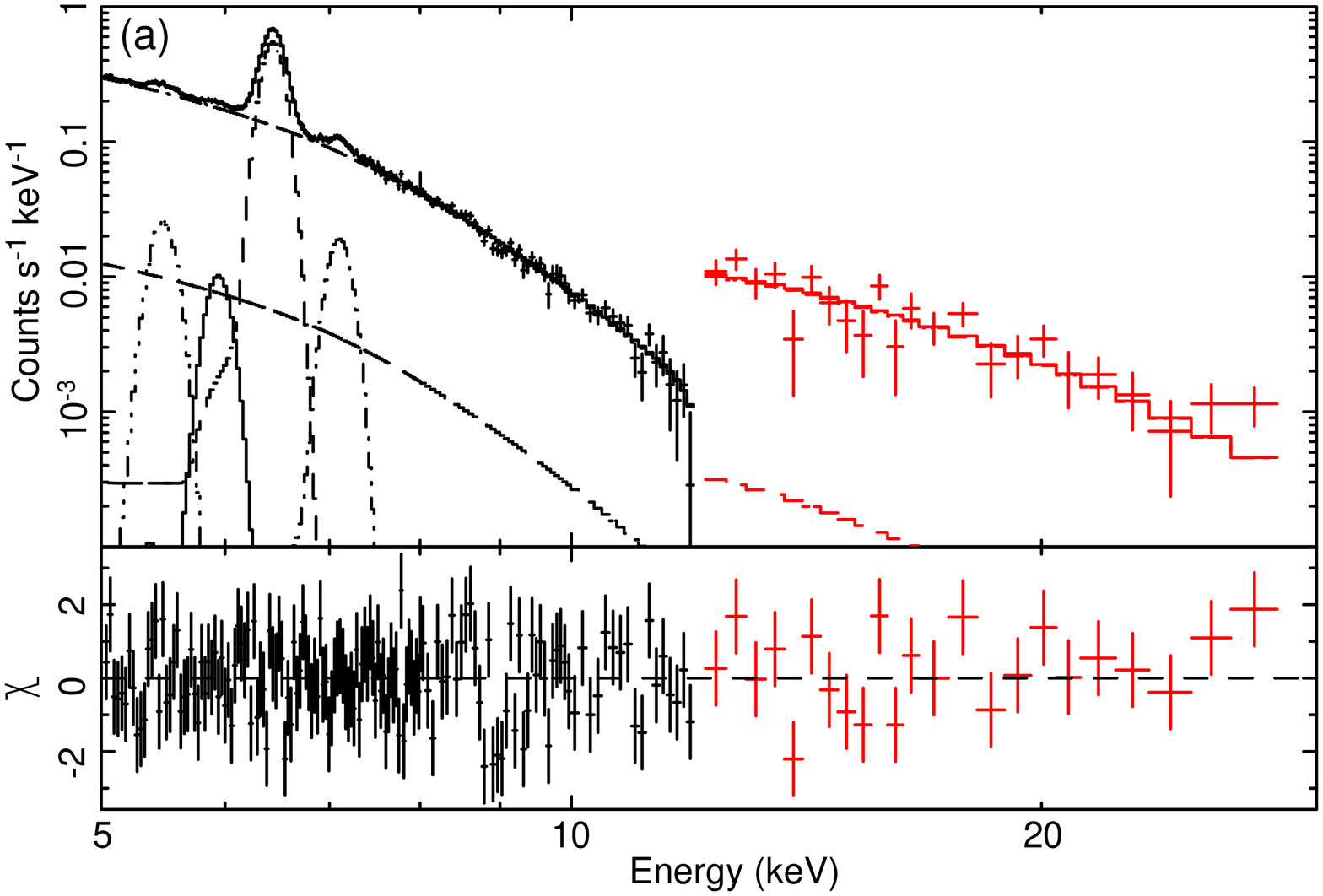}
    \FigureFile(100mm,100mm){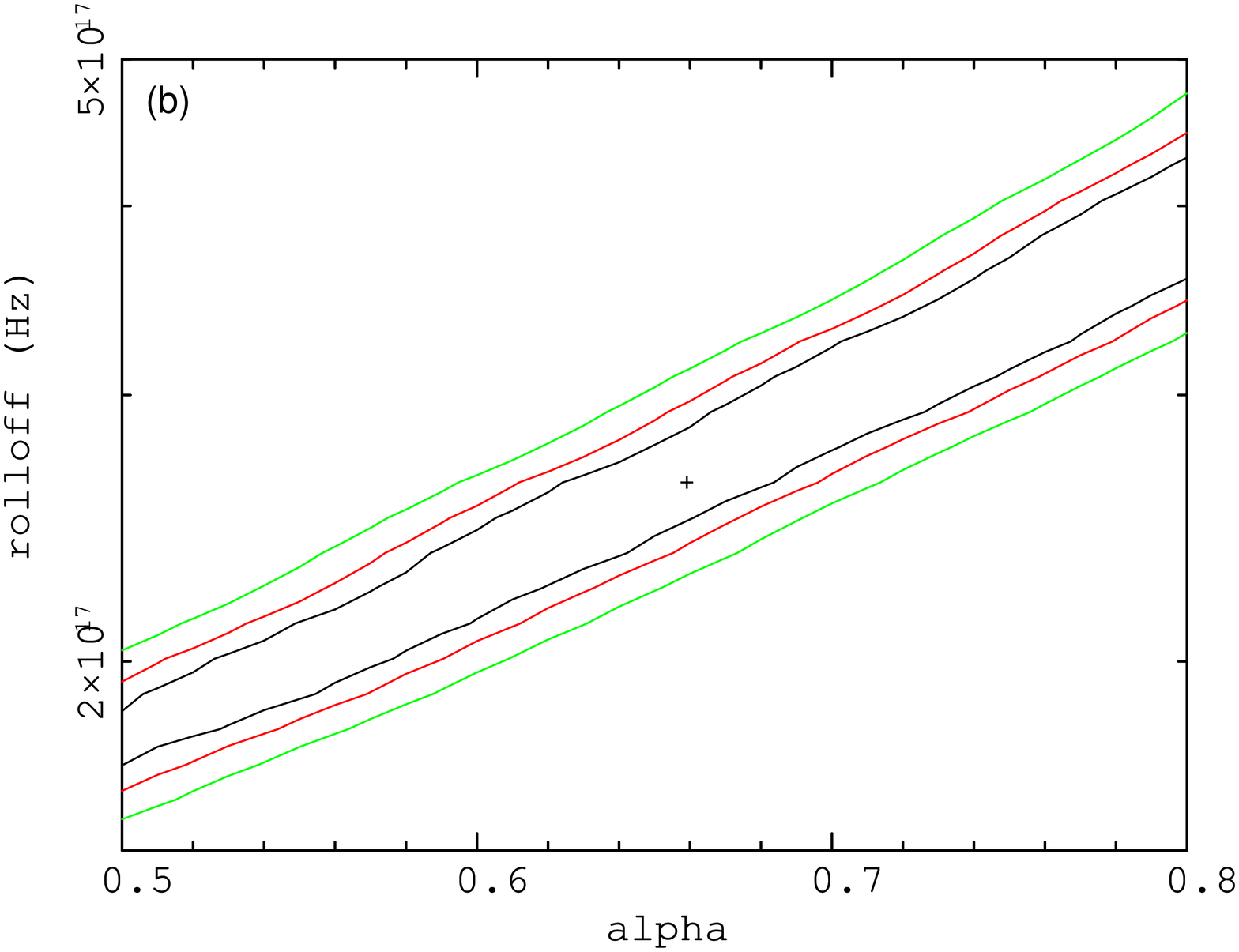}
   \end{center}
   \caption{(a) XIS and PIN spectra fitted with a thermal bremsstrahlung
   and a srcut models. (b) The 68\%, 90\%, and 99\% significance
   contours of the fit.}  \label{fig:xispin_brsr}
  \end{figure}

 The steep slope of the power-law model ($\alpha_p=2.69$) indicates that
 the synchrotron emission we observe in the Suzaku energy region is
 above the roll-off frequency of the spectrum extrapolated from the
 radio regime. Assuming a power-law electron number spectrum $N(E)$ with
 an exponential cutoff energy of
 \begin{eqnarray}
 N(E)=K\ E^{-s} \; \exp(-E/E_{\rm max}),
 \end{eqnarray}
 where $E_{\rm max}$ is the maximum electron energy, we can estimate the
 synchrotron roll-off frequency. Using the electron spectrum, we can
 simply derive the photon spectrum from radio to X-ray by superposing
 the single particle synchrotron emissivities assuming a constant
 magnetic field. This spectral model is implemented in the {\tt XSPEC}
 package as an {\it srcut} model \citep{reynolds1999}. We fitted the XIS
 and PIN spectra with a thermal bremsstrahlung, an {\it srcut} model,
 and Gaussian lines correspond to Cr, Mn, Fe K$_\alpha$, and Fe
 K$_\beta$. In the fits, we fixed the Gaussian parameters and the
 temperature to $kT=4.71$~keV which were derived in
 \S\ref{subsect:broadband_spectra}, while the parameters of the {\it
 srcut} model were left free. As shown in figure \ref{fig:xispin_brsr}a,
 the fit was successfull with $\chi^2$/dof=187.4/173. Although the
 parameters were rather unconstrained as shown in figure
 \ref{fig:xispin_brsr}b, the obtained radio spectral index,
 $\alpha=0.66$, is in a good agreement with the measured value of
 $\alpha=0.65$ \citep{kothes2006}, and the implied radio flux density at
 1~GHz, 45~Jy, is not far from the actual measurement (60~Jy;
 \cite{kothes2006}). Therefore, our {\it srcut} fit to be physically
 appropriate. Then, adopting $\alpha=0.65$, we obtain a roll-off
 frequency of $\nu_{\rm rolloff}=2.6\times10^{17}~{\rm Hz}=1.1$~keV,
 which is slightly higher than that derived by Chandra ($\nu_{\rm
 rolloff}=7.3\times10^{16}~{\rm Hz}=0.3$~keV; \cite{cassam2007}).

 The roll-off frequency $\nu_{\rm rolloff}$ is related to the maximum
 electron energy $E_{\rm max}$ as
 \begin{eqnarray}
  \label{eq:rolloff}
  \rm
  \nu_{rolloff}=5\times10^{15} \; \biggl(\frac{\it B}{10 \; \mu
  G}\biggr) \; \biggl(\frac{{\it E}_{\rm max}}{10 \; TeV}\biggr)^2,
 \end{eqnarray}
 where $B$ represents the magnetic field strength. \citet{warren2005}
 present a summary of magnetic field values and find values in the range
 100 to 400 $\mu$G. Then, equation (\ref{eq:rolloff}) yields $E_{\rm
 max}=$23~TeV with 100 $\mu$G, or 12~TeV for 400 $\mu$~G. Even if we use
 the fiducial magnetic field in our Galaxy, 10~$\mu$G, the maximum
 electron energy is only 72~TeV.

 These maximum electron energies are two to three orders of magnitudes
 below the break energy (``knee'') of the cosmic-ray spectrum around
 1000~TeV. \citet{reynolds1999} pointed out that the $E_{\rm max}$
 values of all Galactic supernova remnants they studied with ASCA data
 were well below the knee energy. To evaluate the role of radiative
 cooling on the electron energy spectrum, we approximate the synchrotron
 radiation loss timescale $\tau_{\rm loss}$ as
 \begin{eqnarray}
  \rm \tau_{loss} (yrs) = 1.2\times10^4 \; \biggl(\frac{\it B}{10 \; \mu
  G}\biggr)^{-2} \biggl(\frac{{\it E}_{max}}{10 \; TeV}\biggr)^{-1}.
 \end{eqnarray}
 With our estimate of {\it E}$_{\rm max}=$23 TeV and magnetic field
 strength of 100 $\mu$G, $\tau_{\rm loss}$ is 52 years. Since the time
 scale is smaller than the age of Tycho's SNR (435 years), the electron
 spectrum should suffer strongly from synchrotron cooling. This has been
 directly detected in the Chandra data of Tycho's SNR as spectral
 steepening in the nonthermal emission across the featureless, thin
 filaments at the rim \citet{cassam2007}.

 \subsection{Abundance of Cr, Mn, and Fe}

 We discovered He-like Ca K$_\beta$, and underionized states of Cr
 K$_\alpha$, Mn K$_\alpha$, and Fe K$_\beta$ lines from Tycho's SNR for
 the first time. The ionization degrees of Cr and Mn were estimated to
 be similar to that of Fe K$_\alpha$ (Ne-like or there-abouts). To
 convert the measured line fluxes or equivalent widths
 (table~\ref{table:lines}) into elemental abundances, a detailed
 emissivity calculation is needed. However, this is beyond the scope of
 the present paper, since the current models that calculate X-ray
 emission under the nonequilibrium ionization conditions do not include
 the species Cr and Mn.

 The relative abundances of trace elements such as Cr and Mn are
 sensitive to the Type Ia supernova explosion mechanism
 \citep{iwamoto1999}, and therefore should provide an important
 diagnostic for these explosions. However, the compositionally
 stratified nature of the supernova ejecta coupled with the inward
 progression of the reverse shock, means that great care needs to be
 taken when comparing observed line fluxes and yields from model
 calculations. In particular, a good fraction of the Fe produced in the
 explosion that Tycho Brahe observed in 1572 still sits unshocked and
 cold in the center of the remnant \citep{badenes2006}. To fully
 interpret the results presented here will require accounting for the
 ejecta structure and its subsequent hydrodynamical evolution to the
 remnant phase. Cr and Mn K$_\alpha$ lines were detected previously only
 from W49B (in which the X-ray emitting gas is nearly in collisional
 ionization equilibrium), where the abundances were found to be
 consistent with solar values \citep{hwang2000, miceli2006}. Detailed
 emissivity calculations for trace species in nonequilibrium hot plasmas
 are strongly encouraged to open this new method for supernova
 nucleosynthesis diagnostics.

 \section{Conclusions}

 We report the first observation of Tycho's SNR by the Suzaku
 satellite. We focused on the X-ray spectra above 4~keV and obtained the
 following results.
 \begin{enumerate}
  \item X-ray emission is securely detected up to at least 30 keV with a
  power-law index of 2.69. When combined with the Chandra spectra at
  Tycho's rim regions, it is natural to interpret that the emission
  comes from synchrotron radiation.
  \item We discovered four emission lines. The line central energies are
  4.56, 5.48, 5.95 and 7.11~keV, corresponding to He-like Ca~K$_\beta$,
  and low ionized Cr~K$_\alpha$, Mn~K$_\alpha$, and Fe~K$_\beta$ blends,
  respectively. When combined with detailed emissivity calculations and
  evolutionary models for the remnant X-ray emission, these new lines
  should provide important new diagnostics to constrain the nature of
  thermonuclear supernovae.
 \end{enumerate}

 \bigskip

 The authors thank all of the Suzaku Science Working Group members for
 their extensive discussions. JPH acknowledges support from NASA grant
 NNG05GP87G.

\newpage
  
  \begin{table}
   \caption{Emission lines above 4~keV
   \footnotemark[$*$].}\label{table:lines}
   \begin{center}
    \begin{tabular}{ccccc}
     \hline
     Line & Central Energy\footnotemark[$\dagger$] & Width &
     Flux\footnotemark[$\ddagger$] &  Equivalent Width \\
     & (keV) & (eV) & $\times$10$^{-5}$ (ph cm$^{-2}$ s$^{-1}$) 
     & (eV) \\
     \hline
     \hline
     Ca K$_\beta$ & 4.56 (4.52, 4.60) & = Fe K$_\alpha$ & 
     1.68 (1.03, 2.16) & 9.7 (1.8, 17.5) \\
     Cr K$_\alpha$ & 5.48 (5.46, 5.50) & = Fe K$_\alpha$ &  
     2.45 (2.03, 2.93) & 23.8 (13.0, 32.3) \\
     Mn K$_\alpha$ & 5.95 (5.90, 6.00) & = Fe K$_\alpha$ &  
     1.13 (0.68, 1.51) & 13.7 (2.7, 27.2) \\
     Fe K$_\alpha$ & 6.445 (6.444, 6.446) & 52.7 (50.0, 54.5) &
     69.1 (68.2, 69.7) & 1.04 (1.02, 1.07) $\times$10$^3$ \\
     Fe K$_\beta$ & 7.11 (7.09, 7.13) & 82.4 (42.1, 122.4) &
     3.55 (2.87, 4.17) & 70.5 (47.1, 95.2) \\
     Ni K$_\alpha$ & 7.51 \footnotemark[$\S$] & = Fe K$_\alpha$ &
     0.82 \footnotemark[$\|$] & 25.1 \footnotemark[$\|$] \\
     \hline
     \multicolumn{5}{@{}l@{}} {\hbox to
     50pt{\parbox{180mm}{\footnotesize
     \vspace{0.2cm}
     \footnotemark[$*$] () denotes 90\% confidence-level error.  \par
     \footnotemark[$\dagger$] Continuum component is a power-law with
     photon index $\alpha_p=2.76\pm0.02$.  \par 
     \footnotemark[$\ddagger$] Assuming annulus source with a radius of
     \timeform{4'.1}.  \par 
     \footnotemark[$\S$] Estimated from the other lines. (see text)\par
     \footnotemark[$\|$] 90\% confidence-level upper limit.
     }\hss}}
    \end{tabular}
   \end{center}
  \end{table}

\newpage
  
\newpage
  
  \begin{figure}
   \begin{center}
    \FigureFile(70mm,70mm){figure03.ps}
   \end{center}
   \caption{PIN spectrum fitted with a power-law model and
   residuals.}\label{fig:pin_spec}
  \end{figure}

\newpage

\newpage

\newpage
  
\newpage
  \begin{figure}
   \begin{center}
    \FigureFile(70mm,70mm){figure07.ps}
   \end{center}
   \caption{The 68\%, 90\%, and 99\% significance contours of two
   thermal bremsstrahlung fit.}  \label{fig:xispin_brbr_cont}
  \end{figure}

\newpage

\end{document}